# A Novel *Bis*-Coumarin Deminished Phosphorylation of Multiple Tyrosine Kinases of Key Signaling Pathways in Melanoma and Inhibits Melanoma Cell Survival, Proliferation, and Migration

**Qurat-ul-Ain[1], Abhijit Basu[3], Sebistein Iben[3], M. Iqbal Choudhary[1, 2], and Karin Scharffetter-Kochanek[3]**

[1]*Dr. Panjwani Center for Molecular Medicine and Drug Research, International Center for Chemical and Biological Sciences, University of Karachi, Karachi-75270, Pakistan*

[2]*H. E. J. Research Institute of Chemistry, International Center for Chemical and Biological Sciences, University of Karachi, Karachi-75270, Pakistan*

[3]*Department of Dermatology and Allergic Disease, Ulm University, 89081Ulm, Germany*

## ABSTRACT

Melanoma is one of the most dangerous skin malignancies due to its high metastatic tendency and high mortality. Activation of key signaling pathways enforcing melanoma progression depends on phosphorylation of tyrosine kinases, and oxidative stress. We here investigated the effect of the new *bis*-coumarin derivative (3,5-DCPBC) on human melanoma cell survival, growth, proliferation, migration, and intracellular redox state, and deciphered associated signal pathways. This novel derivative was found to be toxic for melanoma cells, and non-toxic for their benign counterparts, melanocytes and fibroblasts. 3,5-DCPBC inhibited cell survival, migration and proliferation of different metastatic, and non-metastatic melanoma cell lines through the profound suppression of phosphorylation of the Epidermal Growth Factor receptor, and related downstream pathways. Suppression of phosphorylation of key downstream transcription factors and different tyrosine kinases comprise JAK/STAT, SRC kinases, ERK and MAP kinases (p38alpha), all involved in melanoma progression. Simultaneous and specific targeting of multiple tyrosine kinases and corresponding key genes in melanoma cells makes 3,5-DCPBC a highly interesting anti-melanoma, and anti-metastatic drug candidate which may in the long term hold promise in the therapy of advanced melanoma.

---

*Corresponding Author: Karin.Scharffetter-Kochanek@uniklinik-ulm.de, Tel.: ++(0)731/500-57501, Fax: ++(0)731/500-57502

## 1. Introduction

Malignant melanoma represents most aggressive and deadliest form of skin cancer (Stephanie et al., 2020; Liu, Das, Liu, & Huang, 2018; Prado, Svoboda, & Rigel, 2019). Several systemic therapies (cytotoxic chemotherapy, targeted drugs, immunotherapy, hormonal therapy, radiation therapy, and bio-chemotherapy) have been approved by the US Food and Drug Administration (FDA) (Domingues, Lopes, Soares, & Populo, 2018; Martin & Lo, 2018). Surgery still represents the first treatment option for primary melanoma when metastasis has not yet occurred. At later metastatic stages, systemic therapies are mandatory. The most successful treatment options against non-resettable metastatic melanoma are the immunotherapy with antibodies directed against CTLA-4 and PD-1 (Ollila, Laks, & Hsueh, 2018) or their combination (de Golian, Kwong, Swetter, &

Pugliese, 2016). The overall response rate of 10% to 15% for anti-CTLA-4 (ipilimumab), 25% to 45% for anti-PD-1 (nivolumab or pembrolizumab), and around 60% for their combination were embraced as a recent breakthrough in the therapy of this previously hard-to-treat malignancy. However, due to the resistant nature of malignant melanoma (de Golian et al., 2016), 40% of the patients do not respond to the combined therapy (Lipson et al., 2018). In addition, severe, life-threatening, or fatal side effects (50% only with anti-CTLA-4 immunotherapy, and even more with the combined anti-CTLA-4, and anti-PD-1 therapy), further fuels the urgent quest for new strategies in the battle against metastatic melanoma. Several promising agents have been tested against single protein kinases in malignancies (Aittomaki & Pesu, 2014; de Golian et al., 2016; Eggermont & Robert, 2018; Miklossy, Hilliard, & Turkson, 2013). Phosphorylated-receptors of tyrosine/serine/threonine protein kinases and their downstream targets are molecular players enforcing cell growth, survival, proliferation and migration (Amaral et al., 2017; Cheng, Qi, Paudel, & Zhu, 2011; Day, Sosale, & Lazzara, 2016; Xu & McArthur, 2016). A number of these effectors within distinct signaling pathways are hyperphosphorylated, and, in consequence, hyperactivated in different cancers including melanoma (Maria Mendonça, Luís Soares de, & João Pedro, 2018). Some of the new therapeutic interventions against these kinases have successfully reached clinical trials, while others already have been approved by the US-FDA, and entered clinical routine (Kakadia et al., 2018). Currently, no single systemic therapy has been successful in the long term to prevent melanoma progression. Due to the emergence of resistance against these drugs, multiple phases II, and III melanoma trials are currently underway that are either studying the effect of combination treatments or striving for new synthetic, and natural compounds (Kourie & Klastersky, 2016; Kumar et al., 2017). Combined therapies targeting 2 tyrosine kinases at least delay the development of resistance (Hartmann et al., 2015; Popescu & Anghel, 2017). A newly emerging concept for the treatment of advanced malignant melanoma is based on uncovering new synthetic compounds targeting multiple signaling pathways and their corresponding genes. *Bis*-Coumarins, a benzophenone family of natural compounds, have previously been studied as potential drug candidates against various diseases including different types of cancers. They are, however, not endowed with the required potential to simultaneously target several signaling pathways (Salar et al., 2019). We here set out to evaluate the effect of a newly synthesized derivative of *bis*-coumarins, [3,3'-((3,5-dichlorophenyl) methylene)*bis*(4-hydroxychroman-2-one)] (3,5-DCPBC), on growth, survival, proliferation and migration of non-metastatic and metastatic melanoma cells, and to investigate its underlying molecular mode of action. In the present study, we demonstrate that treatment of human metastatic, and non-metastatic melanoma cells with 3,5-DCPBC profoundly diminished phosphorylation of the Epidermal Growth Factor Receptor (EGFR), and simultaneously suppressed phosphorylation of effectors in key downstream signaling pathways known to enforce melanoma progression. We also uncovered the non-toxic nature of this compound against human melanocytes, the benign counterpart of malignant melanoma cells. In aggregate, we here report for the first time a novel anti-melanoma drug candidate that could not only suppress cellular functions, key for melanoma progression, but simultaneously target multiple phosphor-tyrosine kinases. This is a clinically relevant advancement, and thus, 3,5-DCPBC may hold unique therapeutic potential to be further tested in preclinical, and clinical studies.

## 2. Materials and Methods

### *2.1. Cell lines, reagents, and antibodies*

Metastatic melanoma included A375 (#CRL-1619), SK-Mel-28 (#HTB-72), WM-266-4 (#CRL-1676) melanoma cell lines; non-metastatic melanoma comprised the WM115 (#CRL-1675) melanoma cell line. These melanoma cell lines were procured from A.T.C.C, and normal fibroblasts (FF95) were provided by the Department of Dermatology, University of Ulm, Germany. Human epidermal melanocytes (NHEM, #C-12400), (M2 media, #C-24300) were purchased from Promocell GmbH, Germany. Dulbecco's Modified Eagle's Medium (DMEM #41966161), and F-10 Nutrient Mixture (Ham), #31550] were purchased from Gibco Life Technologies, UK, Amphotericin B # (15290018) was purchased from Gibco Life Technologies,Greenland, Gentamycin # (15710049) were purchased from Gibco, Life Technologies GmbH, China. Pen/strep, phosphate-buffered saline *(*PBS*)*, and L-glutamine solutions were purchased from Biochrome GmbH, Germany. Thiazolyl blue tetrazolium bromide (MTT) dye, #M5655, Accutase solution, (#A6964), human placental type-IV collagen (#C5533), and (DMSO (#276855) were purchased from Sigma-Aldrich Chemie GmbH, Germany. Diff Quick staining solution kit, (#B4132-1A) was purchased from Medion Diagnostic AG, Switzerland. Primary antibodies: anti-BrdU antibody, #555627 was purchased from BD Biosciences, USA. Secondary antibody (Alexa fluor 555 goat anti-mouse, (#A32727) and BrdU dye (#B23151) from Thermofisher Scientific GmbH, Germany. Bradford reagent #5000205 was purchased from Bio-Rad GmbH,USA. Human phospho-kinase array kit #ARY003B from R&D System Germany. Transwell® chambers, #3422 were procured from Corning, NY, USA, chemiluminescence reagent (Signal Fire, #12630S, Cell Signal), chamber slides (Millicell EZ, #C86024 Millipore/Merck GmbH) [3,3'-((3,5-dichlorophenyl) methylene)*bis*(4-hydroxychroman-2-one)] (3,5-DCPBC ) was obtained from the Molecular Bank of Dr. Panjwani Center for Molecular Medicine and Drug Research, International Center for Chemical and Biological Sciences, University of Karachi, Pakistan.

## *2.2. Cell culture*

Metastatic melanoma cell lines used were A375, SK-Mel-28, and WM-266-4; wherease WM115 was used as the non-metastatic melanoma cell line. All melanoma cell lines, and FF95 fibroblasts were grown in DMEM media supplemented with 200 mM L-glutamine, 50 μg/ml gentamycin/ amphotericin B solution, 100 U/mL Pen/Strep, 20% FBS at 37 ºC, and 5% $CO_2$ for 12 hrs prior to assay. Cells were synchronized with starved medium (0.2% F-10 Nut Mix Ham 1X) for further 12 hrs; thereafter, either incubated with 3,5-DCPBC or, in case of non-treated control group, incubated with identical volumes of DMSO in DMEM for the indicated concentrations and incubation times, or they were grown in 20% FBS in DMEM. Human primary melanocytes were grown in M2 media.

## *2.3. Cell cytotoxicity assay*

MTT assay was employed to evaluate cytotoxicity as previously described (Adan, Kiraz, & Baran, 2016). Melanoma cells, and their benign control cells were grown in 96 well plates, as described in secion 2.2. MTT solution was added for 4 hrs and formazan crystals were dissolved in dimethyl sulfoxide for 15 min at room temperature. Absorbance of this solution was recorded at 550 nm to 650 nm, using a microplate reader (Varioskan Lux, Thermofisher Scientific). Cytotoxicity of 3,5-DCPBC was calculated as the relative ratio of optical densities compared to non-treated controls.

## *2. 4. Transwell migration assay*

The Transwell® migration assay was performed as earlier described, with few modifications (Scharffetter-Kochanek et al., 1992). In brief, $1\times10^5$ cells in 200 μL of either starved medium or 20% FBS in DMEM were loaded onto 8-micrometer pore Transwell inserts (upper chambers), and incubated for 2 hrs at 37 °C in 5% $CO_2$. The lower chambers were loaded either with 600 L of 20% FBS in DMEM, chemotactic stimuli (human placental type IV collagen) at a concentration of 100 μg/μL or dissolved in 20% FBS in DMEM, and incubated for 2 hrs at 37 °C in 5% $CO_2$.

After a 4 hrs incubation period, cells were fixed, and stained using the Diff Quik® Stain kit. Non-migrated cells were removed from the upper chambers using cotton swabs. Perforated filters were removed and fixed on OT slides. An average number of migrated cells was counted in five high-power microscopic fields (HPF), randomly chosen at 20X magnification from each of the three technical and four biological replicates. Images of the cells migrated to the downside of the perforated membrane were collected using a Nikon TE300 inverted epifluorescence microscope.

## 2.5. Trypan blue cell viability assay

Melanoma and control cells were incubated with different concentrations of 3,5-DCPBC for 4 hrs, and viable cells were counted using Vi-CEL XR 2.03 (Beckman Coulter, USA) automated cell counter, as described earlier (Kadic, Moniz, Huo, Chi, & Kariv, 2017).

## 2.6. Bromodeoxyuridine (BrdU) incorporation assay

Melanoma and control cells were seeded in 96-well plates, as described in section 2.2 followed by the treatment of 1 µM of 3,5-DCPBC/100 µL of 20% FBS in DMEM in each well for different incubation time. After incubation, 20 ul BrdU/100µL of 20% FBS DMEM was added to each well for another 2 hrs at 37 ºC, followed by fixation of cells with 1:1 acetone solution for 30 min. DNA strains were denatured with 2N HCl, and non-specific binding sites were blocked by 5% BSA dissolved in TBST. 50 µL primary anti-BrdU antibody was suspended in TBST at 1:10 dilution, and added to detect the incorporated BrdU dye. Melanoma cells and control cells in each well were then treated for 2 hrs with 50 µL Alexa Fluor conjugated 555 goat anti-mouse secondary antibody, diluted in TBST at 1:200 dilution. After washing melanoma and control cells with TBST buffer, the fluorescence of mixture in each well was measured in 100 µL of TBST at 548 nm excitation, and 576 nm emission.

## 2.7. BrdU immunofluorescence staining

Cells were grown in Millicell EZ chamber slides coated with poly-l-lysine, and were further incubated with 100 µM of BrdU for 2 hrs. Melanoma cells, and control cells were fixed with 1:1 acetone solution for 30 min, DNA was hydrolyzed with 2N HCl for 30 min and non-specific binding sites were blocked by 5% BSA dissolved in TBST. Melanoma, and control cells were incubated overnight with 500 µL primary anti-BrdU antibody in TBST at a 1:10 dilution, washed thrice with TBST buffer, and thereafter, treated with 500 µL Alexa Fluor conjugated 555 goat anti-mouse secondary antibody suspended in TBST at a 1:200 dilution in each well for 2 hrs at room temperature. Cell nuclei were stained with DAPI (1µL/mL of DAPI in PBS) for 10 min, cells were fixed with formaldehyde and mounted with Fluoromount (Dako). Melanoma and control cells were examined under a fluorescence microscope at 40X magnification with AxioVision A5 microscope (Zeiss Inc., Germany). BrdU positive melanoma, and control cells were counted manually from three independent images. The percentage of proliferating cells was assessed by calculating the number of BrdU-positive cells in the 3,5-DCPBC treated group, and the DMSO treated control group in 5 to 6 random fields.

## 2.8. The human phospho-kinase array assay, and Western blot analysis

The phosphorylation level of kinases was determined with the Proteome Profiler Array Kit (R&D Systems) following the manufacturer's instructions. Protein concentrations were determined by the Bradford protein assay. To block non-specific sites, each of the membranes was incubated in an array blocking buffer for 1 hr. 3,5-DCPBC and DMSO treated cell lysates (334 µL cell lysate/1mL of array buffer corresponding to 200 µg protein lysate) were applied on membranes, and incubated overnight. Thereafter, membranes were washed with 1X washing buffer followed by incubation with 20 mL of the detection antibody for 2 hrs on a shaker at room temperature.

Membranes were thoroughly rinsed with washing buffer thrice, and further incubated with Streptavidin-HRP at room temperature for 30 min. Membranes were washed with 1X washing buffer, for 10 min, and thereafter all membranes were simultaneously exposed to SignalFire plus chemiluminescent reagents for 1 min. Phospho-kinase array data were developed on Vilber FusionFx Chemiluminescence Imager for 1to10 min with multiple exposure times.

### *2.9. Statistical analysis*

Graph Pad Prism software 5 (GraphPad Software, Inc., SanDiego, CA) was used to analyze data. Parametric one-way analysis of variance was used with Tukey's posthoc analysis for comparison between multiple groups of 3,5-DCPBC treated, and DMSO treated groups. Student's t-test (two-tailed) was employed for comparison between the two groups of counted cells taken from representative photomicrographs. Significance was defined as $P < 0.05$. P values were assigned * with $P < 0.01$, ** with $P < 0.001$ and *** with $P < 0.0001$. R studio and Cytoscape were used for data analysis and networking analysis respectively.

## 3. Results

### *3.1. 3,5-DCPBC induced cytotoxicity in metastatic, and non-metastatic melanoma cells*

3,5-DCPBC was synthesized using 2 molecules of 4-hydroxy coumarin and 3,5-dichlorbenzaldehyde in a condensation reaction, as shown in (Figure. 1 A). Cytotoxicity of 3,5-DCPBC was further assessed employing the MTT assay where the activity of NADPH oxidoreductase served as a measure for the extent of cellular toxicity. The results were compared with the optical density determined for non-treated control cells (20% FBS DMEM) or dimethyl sulphoxide (DMSO). We found a concentration and time-dependent increase in cytotoxicity in A375 melanoma cells upon treatment with 3,5-DCPBC ( Figure 1 B and C). 3,5-DCPBC depicts cytotoxicity of more than 50% of melanoma cells upon incubation at concentrations of 1, 10, and 100 µM for 12 hrs (Figure 1 B). Cytotoxicity studies of 3,5-DCPBC on WM-115, and metastatic SK-MEL-28 melanoma cells show similar results (Figure 1 D, E). To explore the specificity of 3,5-DCPBC for its cytotoxicity on melanoma cells, the cytotoxicity of 3,5-DCPBC was further assessed on human melanocytes (NHEM), and fibroblasts (FF95). It was noted, 3,5-DCPBC did not show any cytotoxicity on human melanocytes (NHEM), and fibroblast (FF95) at both time points (Figure 1 G and H).

**Figure. 1**

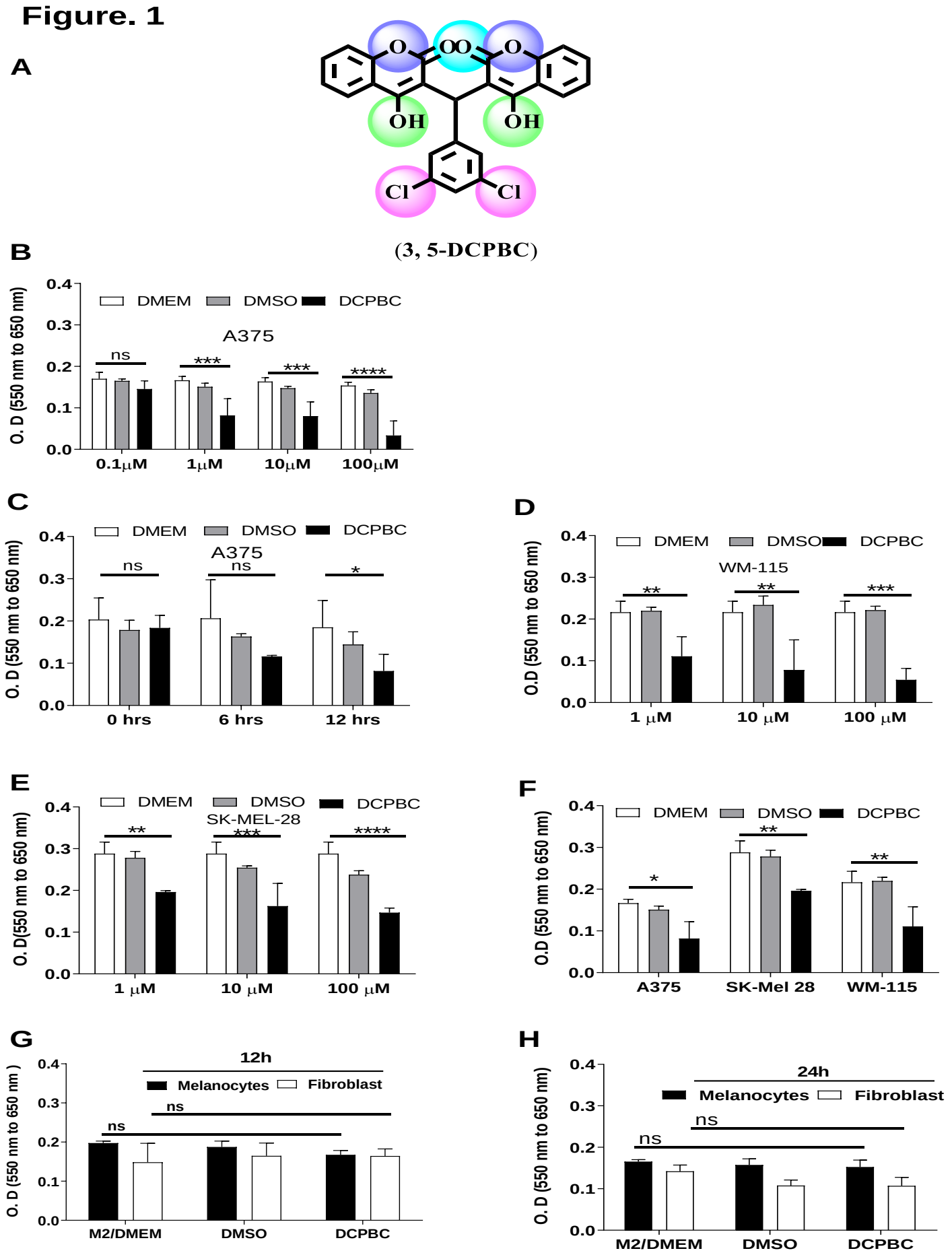

**Figure 1 Structure and IUPAC name of 3,5-DCPBC and its cytotoxic effects on different melanoma cells.** **(A)** Scheme depicting the synthesis of 3,5 DCPBC using 3,5-dichlrobenzaldehyde with 4-hydroxy coumarin in the presence of tetraethylammonium bromide. **(B-H)** Concentration and time-dependent cytotoxic effect of 3,5-DCPBC on metastatic melanoma cells, non-metastatic melanoma cells, primary melanocytes, the benign counterpart of melanoma cells, and fibroblasts. Equal concentration of DMSO dissolved in growth media (M2/DMEM) were used as DMSO controls. Data are presented as the relative ratio of optical densities compared to non-treated controls, (n = 3) ± SEM. SEM, standard error mean. DMSO, Dimethylsulphoxide, M2/DMEM, growth media for melanocytes/ growth media for fibroblasts

### *3.2. 3,5 -DCPBC inhibits migration of melanoma cells*

Since cell migration is a key step in tumor metastasis (Jiang et al., 2015), inhibition of melanoma cell migration is of prime importance. Multi-chamber Transwell® migration assays were employed to assess the effect of 3,5-DCPBC on random, and directed migration of metastatic melanoma (A375, SK-MEL-28) and non-metastatic melanoma (WM-115) cells. The suppressive effect of 3,5-DCPBC was first tested at three different concentrations (0.1, 1, and 10 µM) for 4 hrs on the directed migration of A375 melanoma cells. Of note, 3,5-DCPBC effectively suppressed the directed migration of A375 melanoma cells at all concentrations in a dose-dependent manner (Figure 2 A). Type-IV collagen and 20% FBS served as a strong chemoattractant (positive controls). 3,5-DCPBC impressively inhibited 20% FBS directed A375 melanoma cell migration (Figure 2 A, B). In fact, 3,5-DCPBC suppressed A375 melanoma cell migration by 93.1 % at a concentration of 1 µM compared to the control (Figure 2A, middle panel). Representative photomicrographs of the bottom side of the perforated membranes confirm these results (Figure 2 B). These data imply that 3,5-DCPBC strongly suppresses A375 melanoma cell migration. In order to explore whether the migration suppressing efficacy of 3,5-DCPBC is a more general property, other metastatic and non-metastatic melanoma cells were assessed.

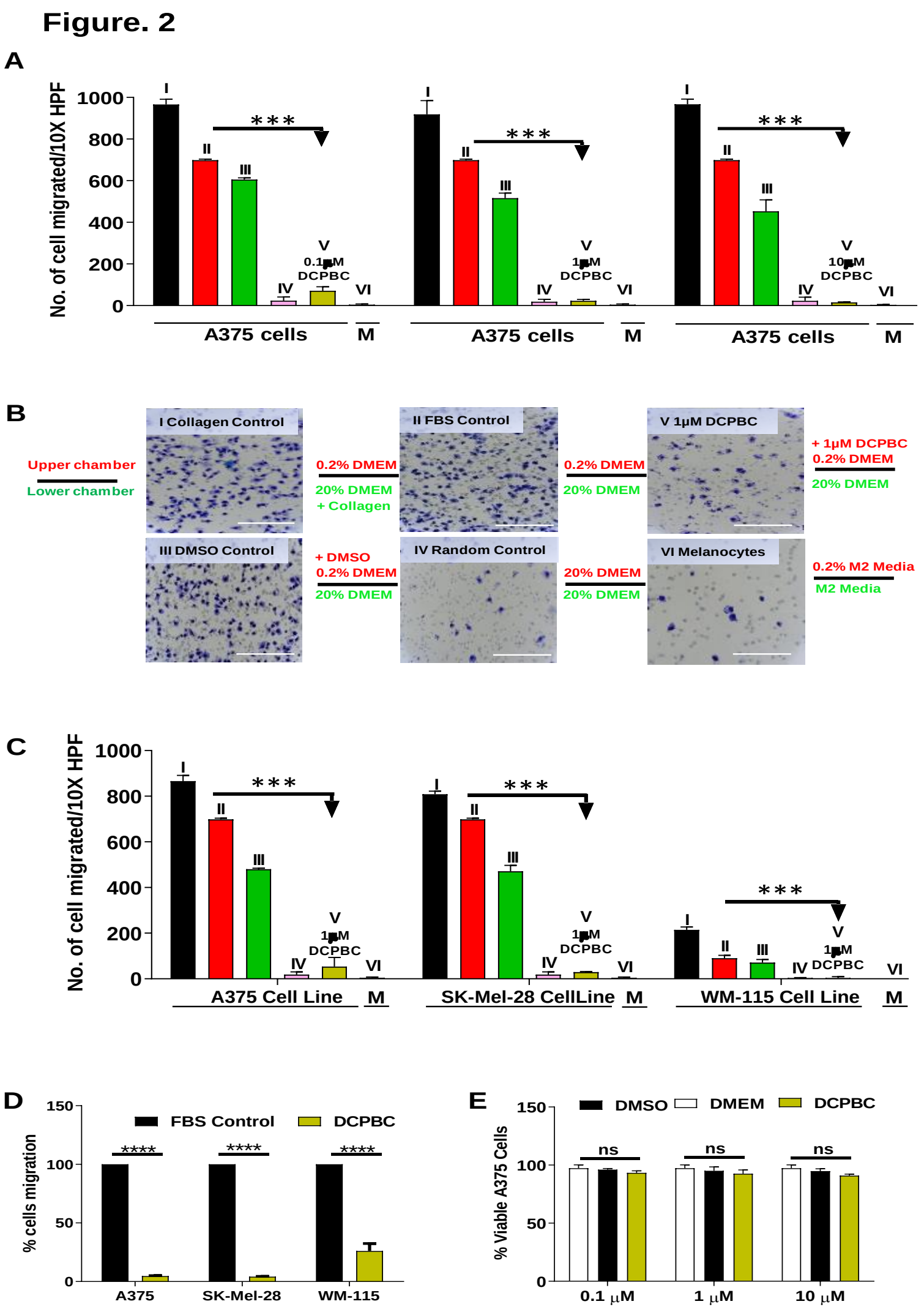

**Figure 2 3,5-DCPBC profoundly inhibits directed melanoma cell migration. (A)** Migration assays were performed as detailed in Materials and Methods. The migration of A375 melanoma cells in the presence of 3,5-DCPBC at the indicated concentrations was tested employing human placental type IV collagen at a concentration of 100 µg/µL dissolved in 600 µL 20% FBS, which serves as a chemoattractant. DMEM, Dulbecco's Modified Eagle's Medium; DMSO; Dimethylsulphoxide; FBS, fetal bovine serum. (I) Positive control (Collagen type-IV), (II) FBS control (0.2% starved FBS/20% FBS DMEM), (III) DMSO Control (DMSO+ 0.2% starved FBS/20% FBS DMEM), (IV) Random Control (20% FBS DMEM/DMEM), (V) (DCPBC) at the indicated concentrations in the upper and lower chambers + (0.2% FBS starved/20% FBS DMEM), (VI) Melanocytes were tested for their migratory response towards 20% FBS in DMEM (chemoattractant) **(B)** Representative photomicrographs of the bottom side of the perforated membranes from the experiments described in Fig. 2 (A); bars, 10 µm. **(C)** Comparison of the inhibitory activity of 1 µM 3,5-DCPBC on the directed migration of different primary and metastatic melanoma cell lines, and melanocytes. The differently tested groups (I-VI) are as described in Fig. 2(A). **(D)** Percentage of inhibition of the directed migration of metastatic, and non-metastatic melanoma cells by 1 µM 3,5-DCPBC. **(E)** Effect of 1 µM 3,5-DCPBC on the viability of metastatic melanoma A375 cells. Data are shown as mean ± S.D; n = 3 replicates; graphs represent one of three independent experiments; ***p < 0.0005 calculated by one-way ANOVA between 3,5-DCPBC treated melanoma cells and controls. Bars represent the means of triplicate determinations, and error bars indicate the standard deviations

Of note, both the directed migration of the metastatic SK-MEL-28 melanoma cells, and the non-metastatic WM-115 melanoma cells were similarly suppressed by 3,5-DCPBC (Figure 2C). Figure 2D depicts the percentage of 3,5-DCPBC suppression of directed melanoma cell migration. This strong inhibitory effect of 3,5-DCPBC on directed melanoma cell migration is a true anti-migratory effect, and not due to a 3,5-DCPBC induced cytotoxicity (Figure 2E). A375 melanoma cells were found to be viable in the range of 90.9 to 93.3% in 3,5-DCPBC treated group as compared to DMSO control groups (97.3%, and 95.03%, respectively) during the studied migration period of 4 hrs.

### *3.3. 3,5 -DCPBC diminishes metastatic melanoma cell proliferation*

The effect of 3,5-DCPBC was further explored on the proliferation of metastatic A375 melanoma cells as a prime event in melanoma progression. For this purpose, BrdU incorporation was studied as previously described (Bundscherer et al., 2008). The DNA thymidine analog 5-bromo-2'-deoxyuridine (BrdU) was allowed to incorporate into rapidly growing metastatic melanoma cells during the S phase of the cell cycle in the presence, and absence of 3,5-DCPBC and thereafter was detected by an antibody directed against BrdU, employing a fluorometer at 548 and 576 nm excitation and emission,

Figure. 3

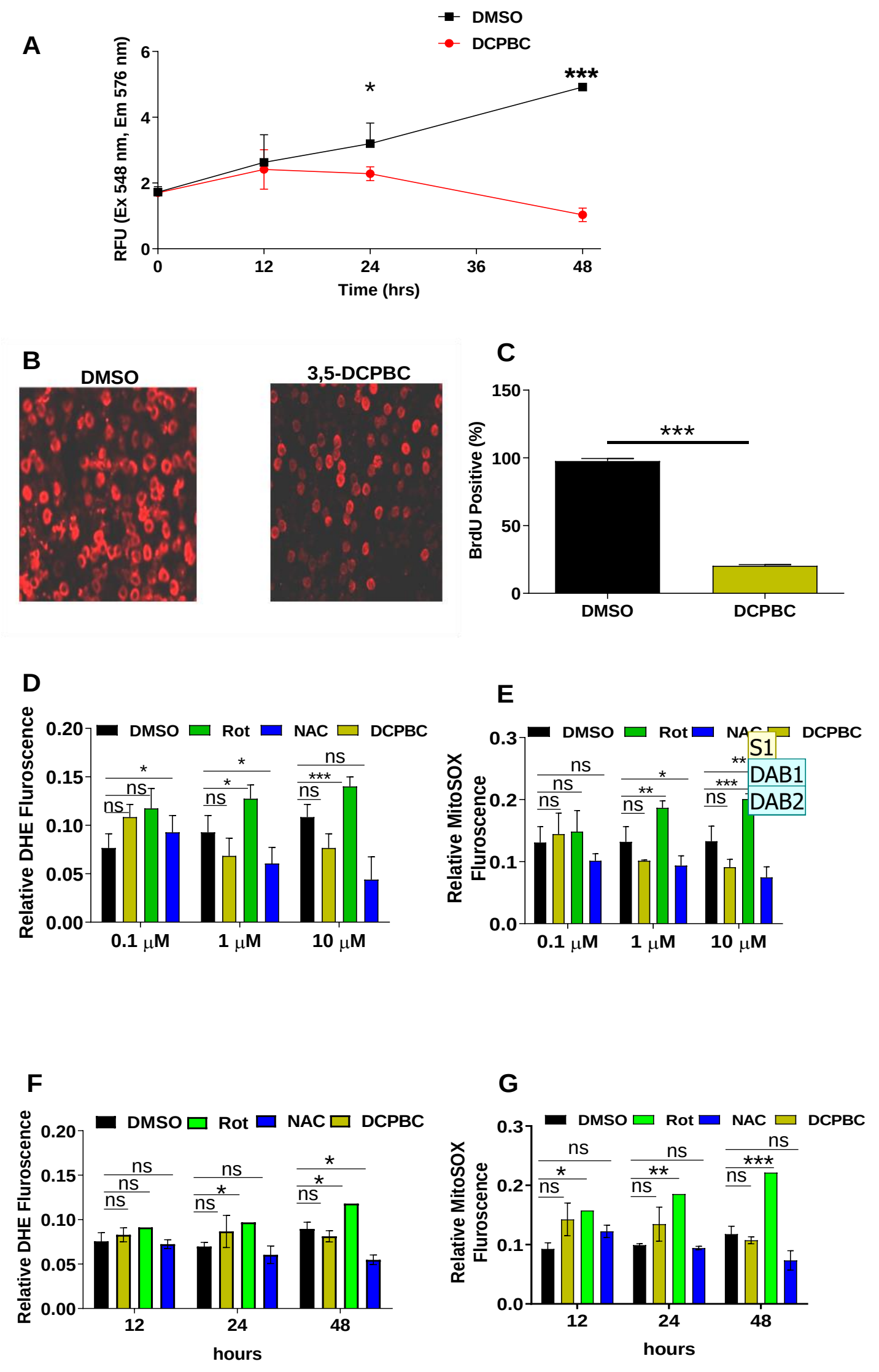

**Figure 3 3,5-DCPBC inhibits melanoma cell proliferation and lacks antioxidant properties.** **(A)** A375 metastatic melanoma cells were cultured in the presence of 1 µM 3,5-DCPBC or 1 µM Dimethyl sulfoxide (DMSO). A pulse of BrdU was provided for 2 hrs. Proliferation of BrdU incorporating A375 melanoma cells was assessed at the indicated time points by a fluorescence spectrometer at excitation 548 nm-emission 576 nm. The red graph corresponds to 3,5-DCPBC treated, and the black graph to non-treated A375 melanoma cells. Data are presented as mean ± standard deviation (SD). *P < 0.05, between treated and non-treated cells at 24 h. ***P < 0.0001 between treated and non-treated cells at 48 h. Independent experiments were repeated three times. DMSO, vehicle control; DMSO, dimethyl sulfoxide. Asterisks indicate statistical difference determined by unpaired student's t-test. **(B)** Fluorescence images of BrdU incorporation indicative of A375 melanoma cell proliferation after treatment with 1µM 3,5-DCPBC (right panel) or, as control with 1µM DMSO (left panel) for 48 hrs. **(C)** Quantification of BrdU positive A375 melanoma cells with original data from Fig. 3 (B). Data are presented as mean± SEM. n> = 3, ***p< 0.0003. Asterisks indicate the statistical difference determined by unpaired student's t-test. **(D)** Dose response relationship between 3,5-DCPBC at the indicated concentrations, and changes in cytosolic ROS levels. *N*-acetylcysteine (NAC) and rotenone (Rot) were used as a ROS quenching and ROS enhancing agents (positive and negative control) respectively. DMSO served as vehicle control. Histograms showing the mean ± SD value of DHE fluorescence intensities indicative of ROS levels. Statistical analysis was carried out by one-way ANOVA. **(E)** The MitoSOX assay was employed to assess dose-dependent changes in the mitochondrial ROS levels, in particular mitochondrial $O_2^{\cdot -}$ upon treatment of A375 melanoma cells with 3,5-DCPBC at the indicated concentrations. Values are shown as mean ± SD (n=3) of fluorescence intensities for ROS. Statistical analysis was carried out by one-way ANOVA. **(F)** Time kinetic of on cytosolic ROS levels upon treatment of A375 melanoma cells with 1 µM 3,5-DCPBC. The fluorescence intensity and statistical analysis was assessed as in Figure 3 (D). **(G)** Time kinetic of mitochondrial ROS levels upon treatment of A375 melanoma cells with 1 µM 3,5-DCPBC. The fluorescence intensity and statistical analysis was assessed as described in Figiure 3 (E)

respectively. 3,5-DCPBC efficiently suppressed melanoma cell proliferation at 48 hrs of treatment (Figure 3A). This anti-proliferative 3,5-DCPBC effect corresponds to an average of the relative fluorescence of 1, compared to 4.9 of the DMSO control at the same time point (Figure 3A). In a complementary approach, BrdU positive cells from immunofluorescence photomicrographs were quantitated, and the percentage of BrdU positive cells were significantly reduced (Figure 3B). These data indicate the anti-proliferative effect of compound 3,5-DCPBC on A375 metastatic melanoma cells.

### *3.4. 3,5 -DCPBC does not alter the intracellular redox state*

Metastatic melanoma progression is known to driven by reactive oxygen species (ROS) be in part, and in particular, the migration and proliferation of metastatic melanoma cells are enhanced by an intracellular increase in superoxide ($O_2^{\cdot -}$) and hydrogen peroxide ($H_2O_2$) levels (Bisevac et al., 2018; Dan Dunn, Alvarez, Zhang, & Soldati, 2015; Denat, Kadekaro, Marrot, Leachman, & Abdel-Malek, 2014).

Figure. 4

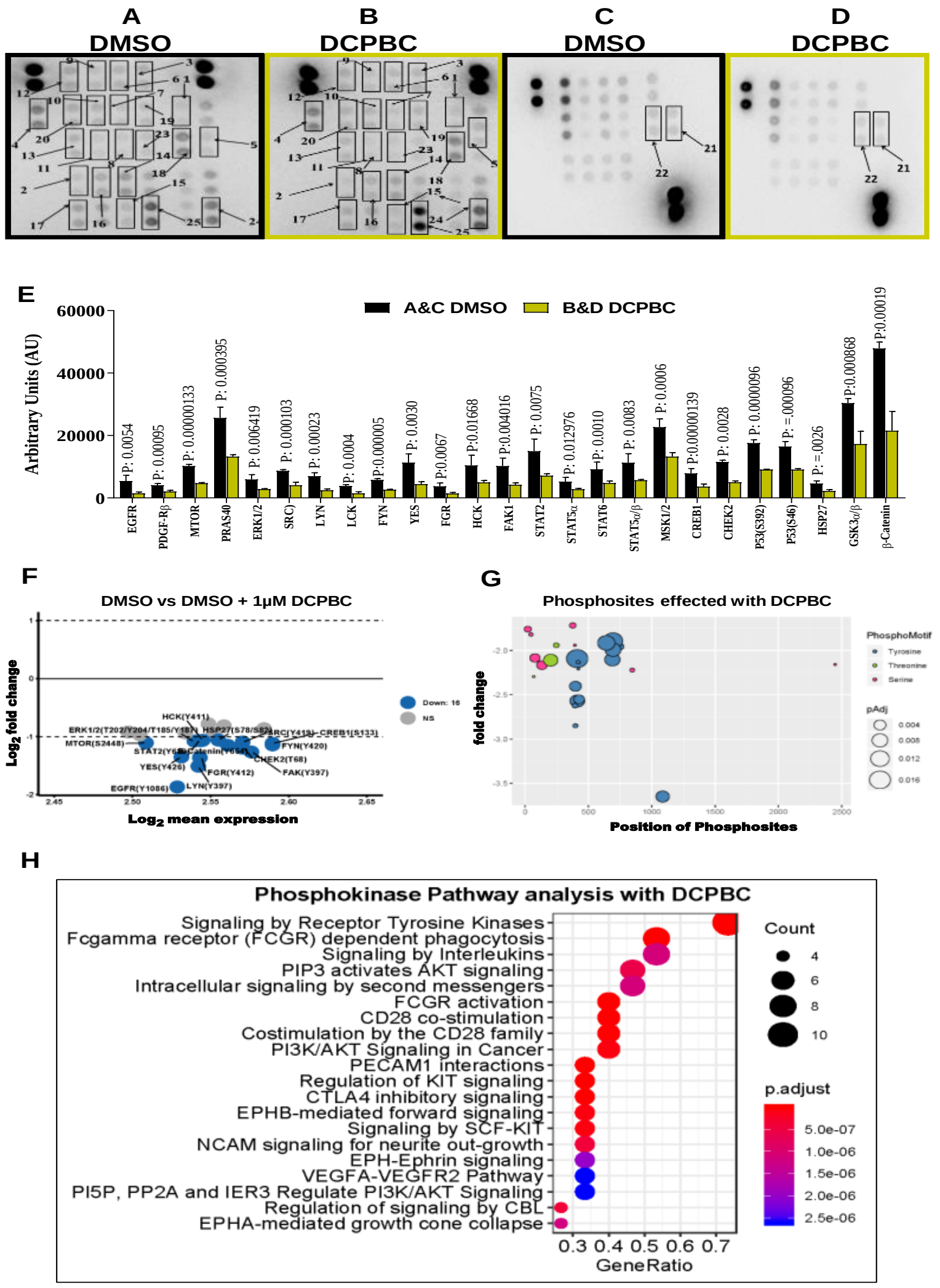

**Figure 4 3,5-DCPBC impacts on phosphorylation of multiple kinases**. Cell lysates from A375 melanoma cells treated with either 1 µM DMSO (control) **(A,C)** or with 1 µM 3,5-DCPBC for 4 hrs **(B,D)** were incubated with PVDF-membranes with anchored antibodies for phosphor-Y1086-EGFR (spot 1), phospho-Y-751-PDGF-β (spot 2), phosphor-S-2448-mTOR (spot 3), phosphor-T246-PRAS40 (spot 4), phospho-T202/Y204/T185/Y187-ERK1/2 (spot 5), phosphor-Y-419-Src (spot 6), phospho-Y-397-Lyn (spot 7), phosphor-Y-394-Lck (spot 8), phosphor-Y-420-Fyn (spot 9), phosphor-Y426-Yes (spot 10), phosphor-Y-412-Fgr (spot 11), phosphor-Y-411-Hck (spot 12), phosphor-Y-397-FAK(spot 13), phosphor-Y689-STAT2 (spot 14), phosphor-Y-694- STAT5α (spot 15), phosphor-Y641-STAT6 (spot 16), phosphor-Y-694/699-STAT5α/β (spot 17), phosphor-S376/S360-MSK1/2 (spot 18), phosphor-S133-CREB (spot 19), phosphor-T68-Chk-2 (spot 20), phosphor-S-392-p53 (spot 21), phosphor-S-46-p53 (spot 22), phosphor-S78/S82-HSP27(spot 23), phosphor-S21/S9-GSKα/β (spot 24), and phosphor-β-catenin (spot 25). Membranes were developed with appropriate secondary antibodies, and the spots were detected using a chemiluminescence based assay, as detailed in Material and Methods. **(E)** Densitometric analysis of the phosphorylation state of multiple kinases from Fig.4 (A-D) employing gel quant software. Results are expressed as "Arbitrary Density units", and presented as mean ± S.D. for 3 independent experiments. Data are analyzed by one-way ANOVA and the statistical difference are shown as *p values. **(F)** Differential phospho-regulation is plotted using MA plot. X-axis represent the mean expression and y-axis represents log2 fold change, the genes that are more than two-fold differentially regulated are shown in blue whereas the non-significant one is shown in grey. **(G)** Phosphorylation motifs were analyzed from the 16 differentially downregulated phosphokinase motifs from (F). **(H)** represents the pathway analysis carried out depicting those genes from the 16 regulated phospho-sites. X-axis represents the gene ratio, whereas the associated pathways are represented in y-axis. The color of the circles represents the level of significance as p values adjusted.

To investigate whether 3,5-DCPBC exerts its anti-proliferative and anti-migratory effect by potential antioxidant properties, the effect of 3,5-DCPBC on mitochondrial and cytosolic superoxide ($O_2^{\bullet -}$), and $H_2O_2$ levels was studied employing the DHE and MitoSOX assays (Dikalov & Harrison, 2014). DHE is a fluorescent probe for the detection of cytoplasmic $O_2^{\bullet -}$ and $H_2O_2$, MitoSOX™ Red identifies mitochondrial $O_2^{\bullet -}$. Briefly, A375 melanoma cells were incubated with MitoSOX™ Red or DHE dye in the presence, and absence of 3,5-DCPBC, and fluorescence was measured by a fluorescence spectrophotometer. High fluorescence indicates low free radical scavenging potential of tested 3,5-DCPBC. 3,5-DCPBC at all tested concentrations did not exhibit any significant antioxidant effect on cytosolic or mitochondrial $O2^{\bullet -}$ and $H_2O_2$ (Figure 3 D, E) in A375 melanoma cells. By contrast with 3,5-DCPBC, both rotenone, a complex I inhibiting agent which increases mitochondrial $O_2^{\bullet -}$ as well as *N*-acetyl cysteine (NAC) serving as a substrate for the $H_2O_2$ removing classical glutathione peroxidase, depict strong effects on ROS levels (Figure 3D, E, F, G). We found that 3,5-DCPBC did not modulate intracellular redox state by scavenging mitochondrial/cytosolic $O_2^{\bullet -}$ or $H_2O_2$ concentrations. These data imply that the observed suppression of proliferation, and migration on melanoma cells is not due to antioxidant properties of 3,5-DCPBC.

Figure. 5

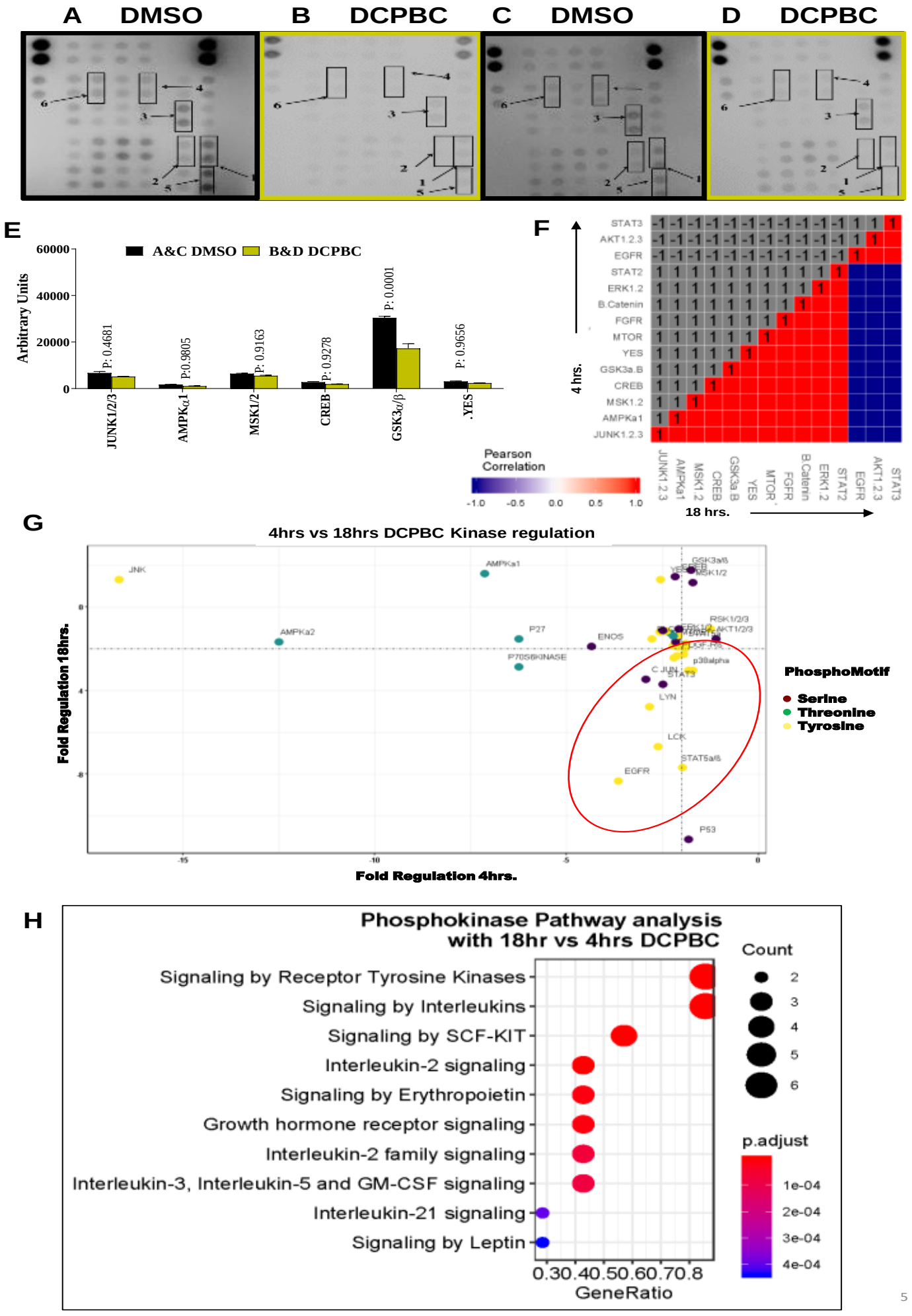

**Figure 5 3,5-DCPBC reduces phosphorylation (activation) of multiple kinases in melanoma cells. (A, B, C, D)** A375 metastatic melanoma cells were treated with 1 µM 3,5-DCPBC for 18 hrs. Cell lysates from the DMSO treated (control) (A), and 3,5-DCPBC treated (B) A375 melanoma cells were incubated with PVDF-membranes with anchored antibodies for phosphor-T183,221/Y185,223-JUNK1/2/3 (spot 1), phosphor-T183-AMPKα1(spot 2), phosphor-S376/S360-MSK1/2 (spot 3), phosphor-S133–CREB (spot 4), phosphor-S21/S9-GSK-3αβ/ (spot 5), and phosphor-Y420-Yes (spot 6). Membranes were developed with appropriate secondary antibodies, and spots were detected using a chemiluminescence based assay, as described in Material and Methods. **(E)** The phosphorylation (activation) state of kinases was determined by densitometry analysis from data of Fig. 5 (A, B, C, D). Results are expressed as “Arbitrary Density units”, and presented mean ± S.D. for 3 independent experiments. Data is analyzed by one-way ANOVA and the statistical difference are shown as *p values. **(F)** The association between x-axis 4 hrs downregulation vs y-axis 18 hrs down-regulation (blue colored box) includes AKT1, STAT3, EGFR. **(G)** Correlation plot between highly correlated genes at 4 and 18 hrs (F) is depicted as scatterplot, indicting differential phosphorylation between 375 melanoma cells treated with 3,5-DCPBC for 4 vs 18 hrs. The x-axis represents fold change of 4hrs DCPBC/DMSO treatment, and the y-axis represents fold change of 18hrs DCPBC/DMSO treatment. **(H)** Pathway analysis of significantly down-regulated genes in 375 melanoma cells treated with 3,5-DCPBC for 4 vs 18 hrs. Differentially regulated genes with more than 2-fold change was included in this pathway analysis. The pathways are represented in the y-axis and the gene ratio in the x-axis. Color represents the level of significance, and p value adjusted. Data is analyzed by one-way ANOVA and the statistical difference are shown as *p<0.05, **p<0.05 and ***p<0.001 when compared between treated and control cells

## *3.5 Molecular mechanism of action of 3,5-DCPBC*

To gain insight into the mechanism of how 3,5-DCPBC inhibits melanoma cell survival, growth, proliferation, and migration, a phospho-proteome profiling array was employed. A375 melanoma cells were treated with 3,5-DCPBC at a concentration of 1 µM for 4, and 18 hrs. For each time point, a total of 43 kinase phosphorylation sites and 2 related proteins (HSP, and tumor suppressor proteins) were analyzed. A significant effect of compound 3,5-DCPBC was observed on 31 phosphorylation sites of 25 different kinases at 4 hrs, and 11 phosphorylation sites of 6 different kinases at 18 hrs. Altogether, 36 phosphorylation sites of 27 different kinases were found to be potential targets of 3,5-DCPBC as opposed to the DMSO control, (Figure 4 A, C, and B, D). A list of non-affected kinases is attached as (Supplementary Tables 1 and 2). Densitometric analysis of significant phosphorylation sites of the 25 kinases along with their associated P values is shown in Figure 4 E. Sixteen different kinases were more than 2-fold down-regulated upon treatment with 3,5-DCPBC as compared to DMSO (Figure 4 F). The MA plot depicts $Log_2$ fold changes on y-axis, and $Log_2$ mean expression on x-axis (Figure 4 F). Eleven kinases did not show any significant changes. Among the 16 differentially down-regulated phosphokinases, predominantly members of the family of the tyrosine kinase family were suppressed,

**Figure. 6**

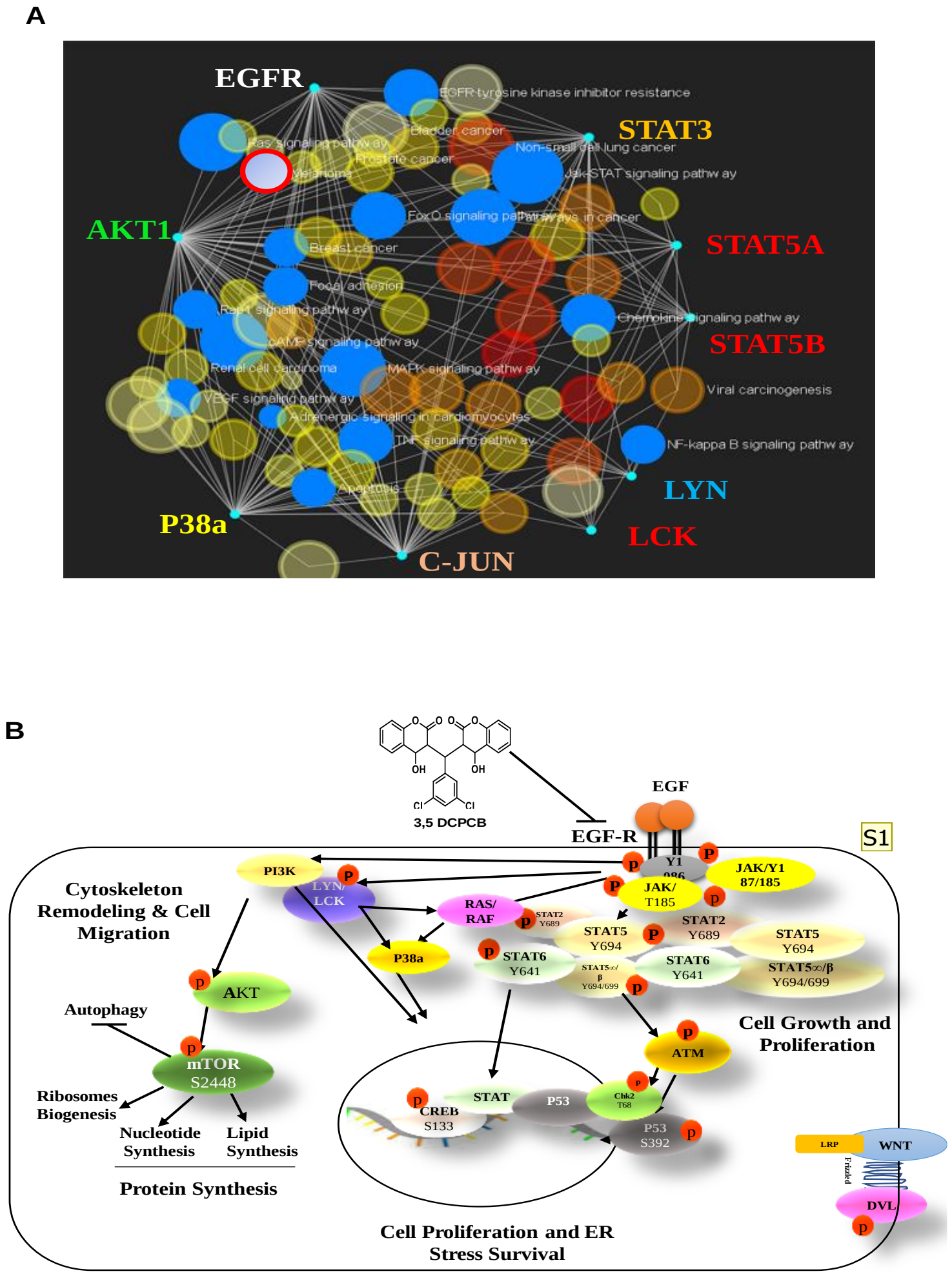

**Figure 6 3,5-DCPBC impacts on important regulatory hubs in melanoma progression. (A)** Gene enrichment and networking depicts relevant nodes and edges, among them 9 genes, namely, EGFR, STAT3, STAT5A, STAT5B, AKT1, LYN, LCK, P53, and P38a. Nodes and the edges are subdivided into signaling pathway (represented in blue color), some of them are associated with a variety of cancers including melanoma (the latter highlighted in red). All uncovered 9 nodes show a strong interdependence with the key signaling pathways [EGFR/STAT axis] / [EGFR/SRC axis]/ [EGFR/RAS/MAPK axis]/ [EGFR/PI3K/AKT axis]. The nodes were unweighted and connected using Waserman and Faust algorithm. Network mapping shows that these genes depict a strong cross talk, and are associated with malignancies including malignant melanoma. **(B)** Summary scheme depicting 3,5-DCPBC targeting EGFR and SRC kinases, AKT/mTOR and RAS/RAF/MAP kinases. Taken together, 3,5-DCPBC shows excellent tyrosine kinase inhibiting properties. Most excitingly, it is endowed with the potential to target multiple kinases to prevent cellular migration (via SRC), ER stress survival (via p38α), proliferation and protein synthesis (via AKT/mTOR). Red circles with p indicate the genes in the signaling cascade, which could be post translationally regulated with phosphorylation at kinase sites as well as which were found to be significantly

phosphorylation of serine or threonine kinases was less frequently affected (Figure 4 G). The y-axis represents the fold change, and the x-axis represents the position of the phospho-sites in the gene amino acid sequence (Figure 4 G).

6

The tyrosine kinase phospho-motif (blue filled circles) is the most affected phospho-motif in A375 melanoma cells treated with 3,5-DCPBC as compared to threonine (green filled circles) or serine (pink filled circles). Phospho-motif mapping thus allowed us to uncover major kinases and signaling pathways targeting key genes for melanoma progression. Pathway analysis was further employed for the most suppressed kinases with downregulation of phospho-motifs to explore which signaling pathways are most prominently affected (Figure 4 H). The x-axis represents the gene ratio which is presented as % of total differential gene expression (DGE) in all GO clusters, and the y-axis represents the associated pathway. The color refers to the P-value, and the count as the number of occurrences of this gene per GO cluster. Pathway analysis with the differentially down-regulated phosphor-motifs depict receptor tyrosine kinases, inflammatory interleukins and ERK1/2, AKT, and mTOR as the major signaling cascades that are affected after treatment of melanoma cells with 3,5-DCPBC. Briefly, phosphor tyrosine modulation at a specific residue of Epidermal Growth Factor Receptor (EGFR$^{Y1086}$) and Fc gamma Receptors$^{Y412}$, and their downstream target JAK/STAT pathway were the most suppressed target including STAT2$^{Y689}$, STAT5α$^{Y697}$, STAT5α/β$^{Y694/Y699}$, and STAT6$^{Y641}$. Among the other 3,5-DCPBC suppressed pathways highly important for melanoma progression, are mTOR$^{S2448}$, ERK1/2 $^{Y204}$/ $^{Y187}$, SRC$^{Y419}$, and β-Catenin$^{Y654}$. In order to understand whether inhibition of phosphosites following treatment of 375 melanoma cells with 3,5-DCPBC will persist for 18 hrs, phosphokinase array analysis was performed (Figure 5A, B, C, D): Densitometric analysis from A-D depicts the downregulation of phosphor-sites is not maintained in all identified phosphokinases (Figure 5 E). Of note, as shown in the blue box of the correlation plot (**Fig. 5F**), EFGFR, AKT1, 2, and 3 and STAT3 at 18 hrs of 3,5 DCPBC are still profoundly suppressed. These phosposites are mainly from tyrosine kinases (Figure 5G), and they impact on the genes involved in tyrosine kinase signaling (Figure 5H).

## 4. Discussion

The major unprecedented finding of this study is that the newly synthesized *bis*-coumarin derivative referred to as 3,5-DCPBC has profound inhibitory properties on key steps of malignant melanoma progression (Wirbel, Cutillas, & Saez-Rodriguez, 2018). Accordingly, proliferation, migration, and survival of melanoma cells - by contrast to melanocytes, their benign counterpart, or fibroblasts - are impressively downregulated *in vitro* in the presence of moderate concentrations of 3,5- DCPBC. The strong effect on key features of melanoma progression is not due to any antioxidant property of 3,5-DCPBC, though ROS have been reported to be associated with melanoma pathophysiology, but rather a consequence of its outstanding suppressive potential of the major different tyrosine phosphokinases involved in melanoma progression (see summary graph Figure. 6). In fact, phosphokinases modulate several cellular functions (Nishi, Shaytan, & Panchenko, 2014), and activation of multiple phosphokinases has been implicated in melanoma, and many cancers including autophosphorylation of Epidermal Growth Factor Receptor (EGFR) (Boone et al., 2011; Davies, 2012) Of note, compound 3,5-DCPBC significantly diminished the phosphorylation of the $Y^{1086}$ of EGFR and $Y^{751}$ of PDGFR at 4 hrs (Figure 4 E, F, Figure 5E and Supplementary Table-1). As phosphorylation of these residues of EGFR and PDGFR regulates various signal transduction pathways (mTOR, SFK, JAK-STAT, and MAPK) involved in cell proliferation, cell migration as well as cell survival, are found to be hyperphosphorylated in various cancers including melanomas (Dratkiewicz, Simiczyjew, Pietraszek-Gremplewicz, Mazurkiewicz, & Nowak, 2019; Girotti et al., 2013; Nazarian et al., 2010; Welsh, Rizos, Scolyer, & Long, 2016), we further studied the downstream pathways, and their associated genes that are regulated through $EGFR^{Y1086}$ phosphorylation ( Figure 4 H). Hyperphosphorylation of mTOR at its threonine $Thr^{2446}$, serine residues ($Ser^{2448}$, and $Ser^{2481}$) *via* EGFR- ERK–S6K1 and PI3K/AKT axis (Copp, Manning, & Hunter, 2009), linked to growth in various types of cancers and melanoma (Holroyd & Michie, 2018), was markedly downregulated upon 3,5-DCPBC treatment, as confirmed by downregulation of 16 genes within the mTOR pathway (Figure 4 F). Suppression of phosphorylation (activation) of the mTOR pathway (mTOR, PRAS40, and ERK1/2/3) was observed only at 4 hrs after 3,5-DCPBC treatment. EGFR phosphorylation can activate the Src family of kinases (SFKs) (Halaban et al., 2019), they consequently play key roles in cell differentiation, motility, proliferation, and survival (Irwin, Bohin, & Boerner, 2011), Tyrosine phosphorylation of members of SFKs including $SRC^{Y419}$, $LYN^{Y397}$, $LCK^{Y394}$, $FYNY^{420}$, $Yes^{Y426}$, $FGR^{Y^{412}}$, $HCK^{Y41,}$ and $FAK^{Y319}$ were all profoundly down-regulated in A375 malignant melanoma cells treated with 3,5-DCPBC. The finding that the newly synthesized compound 3,5-DCPBC has the potential to simultaneously suppress many phospho-tyrosine kinases, among them EGFR and SRC, is of major clinical relevance. In fact, the simultaneous inhibition of tyrosine phosphorylation of EGFR and SRC kinases has recently been reported to overcome the frequently developing BRAF resistance in melanoma (Girotti et al., 2013). Of major importance among other contributing downstream signaling pathways of EGFR is the JAK/STAT axis (Pansky et al., 2000), whose tyrosine phosphorylation at $STAT2^{Y689}$, $STAT5\alpha^{Y697}$, $STAT5\alpha/\beta^{Y694/Y699,}$ and $STAT6^{Y641}$ is significantly reduced in A375 malignant melanoma cells when treated with 3,5-DCPBC. In addition, phosphorylation on $Y^{705/727}$ and $Y^{694/699}$ of STAT3, and STAT5α/β most likely in cooperation with MSK (nuclear serine/threonine protein kinases that are activated by members of MAPK families), enforce phosphorylation of ERK1/2 (extracellular signal-regulated kinase) or of p38 in a growth factor receptor-independent signaling pathways (Koul, Pal, & Koul, 2013). These pathways are essential for melanoma progression, and their inhibition by 3,5-DCPBC and may suppress melanoma cell survival and proliferation (Estrada, Dong, & Ossowski, 2009; Mirmohammadsadegh et al., 2007; Zhuang et al., 2005)

## Conclusion

In conclusion, with urgent need for new therapeutics to target multiple tyrosine kinases as a newly emerging concept for advanced melanoma, 3,5-DCPBC might hold promise for an efficient alternative approach to currently established therapies (Girotti et al., 2013; Hartmann et al., 2015; Kourie & Klastersky, 2016; Vojnic et al., 2019). In fact, we here discovered that 3,5-DCPBC is highly suppressive on many steps and pathways of melanoma cell progression at very low concentrations, and is non-toxic for non-tumorous melanocytes, and fibroblasts. The efficient combined targeting of EGFR, SRC, and MAPK by 3,5-DCPBC in melanoma cells may assist clinicians in long term prevention of tumor cell proliferation, and even overcoming drug resistance. Further preclinical data are now required, and, thereafter clinical trials for this compound are the next highly interesting steps.

## Acknowledgments

We thank all colleagues of the Department of Dermatology and Allergy, University of Ulm, Germany for many helpful discussions. We are especially thankful to Dr. Meinhard Wlaschek, Marius Alupi, Filipa Ferreira, Karmveer Singh, Saira Munir, and Adelheid Hainzl for their kind support and to Priv.-Doz. Dr. med, Sebastian Iben for providing technical support for the proteome profiling array assay.

## Conflict of interests

The authors declare no conflicts of interest.

## Funding

This work was financially supported by The International Union of Biochemistry and Molecular Biology (IUBMB) with a stipend for Q.-A. The project was financially supported by the Department of Dermatology and Allergic Disease, Ulm University, Germany, and the Petron in Chief, ICCBS, Prof. Dr. Atta-ur-Rahman.

## Author Contribution

Qurat-ul-Ain contributed to study design, manuscript writing, performed all cell and molecular biology experiments, data analysis, and statistical analysis. Sebistain Iben helped to perform proteome profiling array assay. Abhijit Basu largely contributed to project design, technical support to all experiments, statistical and bioinformatics analysis, and correction of the manuscript. M. Iqbal Choudhary and Karin Scharffetter-Kochanek supervised the project, reviewed and corrected the manuscript.

## Supplimentry Material

Table 1. Non affected kinases of proteome profiling assay at 4 hrs.

| Protein Kinases | Phosphorylation Site | Statistical Analysis<br>* ρ ≤0.05<br>** ρ≤0.005<br>***ρ≤0.0005 |
|---|---|---|
| **p38alpha** | T180/Y182 | 0.33 |
| **JUNK1/2/3** | T183/Y187/T185/Y185 /221/223 | 0.06 |
| **AMPKα1** | T183 | 0.14 |
| **AKT1/2/3** | S473 | 0.09 |
| **AMPKα2** | T172 | 0.08 |
| **STAT5β** | Y699 | 0.06 |
| **AKT1/2/3** | T308 | 0.66 |
| **P70S6KINASE** | T389 | 0.35 |
| **P53** | S15 | 0.09 |
| **C JUN** | S63 | 0.29 |
| **P70S6KINASE** | T421/S424 | 0.09 |
| **RSK1/2/3** | S380/S386/S377 | 0.96 |
| **ENOS** | S1177 | 0.53 |
| **STAT3** | Y705 | 0.27 |
| **P27** | T198 | 0.66 |
| **PLCγ1** | Y783 | 0.66 |
| **STAT3** | S727 | 0.33 |
| **WNK1** | T60 | 0.72 |
| **PYK2** | Y402 | 0.86 |
| **SHP60** | | 0.89 |

**Abbreviations; Threonine: T, Tyrosine: Y, Serine: S.**

Supplementary Table 1. Table shows the list of genes (left column) and various phosphorylation sites that were associated with transcription factors (middle column) and the ratio of the phosphorylation levels of A375 MM cells treated with 3,5-DCPBC vs DMSO in arbitrary units for a period of 4 hrs (right column).

Table 2. Non affected kinases of proteome profiling assay at 18 hrs.

| **Protein Kinases** | **Phosphorylation Sites** | **Statistical Analysis** <br> *** ρ ≤0.05** <br> **** ρ≤0.005** <br> *****ρ≤0.0005** |
|---|---|---|
| **p38alpha** | T180/Y182 | 0.58 |
| **ERK1/2** | T202/T185/Y187/Y204 | 0.70 |
| **EGFR** | Y1086 | 0.12 |
| **AKT1/2/3** | S473 | 0.90 |
| **mTOR** | S2448 | 0.60 |
| **SHP27** | - | 0.96 |
| **AMPKα2** | T172 | 0.60 |
| **β -Catenin** | - | 0.54 |
| **Src** | Y419 | 0.43 |
| **Lyn** | Y397 | 0.21 |
| **Lyk** | Y394 | 0.15 |
| **STAT2** | Y689 | 0.74 |
| **STAT5α** | Y694 | 0.57 |
| **Fyn** | Y420 | 0.41 |
| **Fgr** | Y412 | 0.84 |
| **STAT6** | Y641 | 0.51 |
| **Hck** | Y411 | 0.52 |
| **STAT5α/β** | Y694,Y699 | 0.13 |
| **FAK** | Y397 | 0.87 |
| **Chk-2** | T68 | 0.80 |
| **PDGFB** | Y751 | 0.44 |
| **STAT5α/β** | Y694,Y699 | 0.97 |
| **PRAS40** | T246 | 0.13 |
| **P53** | S392 | 0.86 |
| **P53** | S15 | 0.91 |
| **P53** | S46 | 0.46 |
| **AKT1/2/3** | T308 | 0.76 |
| **P70S6KINASE** | T389 | 0.32 |
| **C JUN** | S63 | 0.34 |
| **P70S6KINASE** | T421/S424 | 0.16 |
| **RSK1/2/3** | S380/S386/S377 | 0.80 |
| **ENOS** | S1177 | 0.23 |
| **STAT3** | Y705 | 0.40 |
| **P27** | T198 | 0.16 |
| **PLCγ1** | Y783 | 0.36 |
| **STAT3** | S727 | 0.55 |
| **WNK1** | T60 | 0.45 |
| **PYK2** | Y402 | 0.41 |
| **SHP60** | - | 0.40 |

Supplementary Table 2. Table shows the list of the genes (left column) and various phosphorylation sites that were associated with transcription factors (middle column) and the ratio of the phosphorylation levels of A375 MM cells treated with 3,5-DCPBC vs DMSO in arbitrary units for a period of 18 hrs (right column).